\documentclass[twocolumn, amsmath, amssymb, aps, nofootinbib, longbibliography]{revtex4-2}

\usepackage{amssymb}
\usepackage{amsmath}
\usepackage{mathtools}
\usepackage{bm}
\usepackage{graphicx}

\graphicspath{{figure/}}

\DeclareMathOperator{\erf}{erf} 
\DeclareMathOperator{\sgn}{sgn} 

\newcommand{\Deltad}{\Delta_\mathrm{d}} %
\newcommand{\Deltac}{\Delta_\mathrm{c}} %

\usepackage[dvipsnames]{xcolor}

\begin{document}

\title{Resolution of similar patterns in a solvable model of unsupervised deep learning with structured data}

\author{Andrea Baroffio}
\affiliation{Universit\`a degli Studi di Milano, via Celoria 16, 20133 Milano, Italy}

\author{Pietro Rotondo}
\affiliation{Universit\`a degli Studi di Milano, via Celoria 16, 20133 Milano, Italy}

\author{Marco Gherardi}\email[]{marco.gherardi@unimi.it}
\affiliation{Universit\`a degli Studi di Milano, via Celoria 16, 20133 Milano, Italy}
\affiliation{Istituto Nazionale di Fisica Nucleare, Sezione di Milano}

\begin{abstract}
Empirical data, on which deep learning relies, has substantial internal structure, 
yet prevailing theories often disregard this aspect.
Recent research has led to the definition of structured data ensembles,
aimed at equipping established theoretical frameworks with interpretable structural elements,
a pursuit that aligns with the broader objectives of spin glass theory.
We consider a one-parameter structured ensemble where data consists of correlated pairs of patterns,
and a simplified model of unsupervised learning, whereby the internal representation
of the training set is fixed at each layer.
A mean field solution of the model identifies
a set of layer-wise recurrence equations for the overlaps between 
the internal representations of an unseen input and of the training set.
The bifurcation diagram of this discrete-time dynamics is topologically inequivalent
to the unstructured one, and displays transitions between different phases,
selected by varying the load 
(the number of training pairs divided by the width of the network).
The network's ability to resolve different patterns
undergoes a discontinuous transition to a phase where
signal processing along the layers dissipates differential information about an input's proximity to
the different patterns in a pair.
A critical value of the parameter tuning the correlations separates regimes where
data structure improves or hampers the identification of a given pair of patterns.
\end{abstract}

\maketitle

\section{Introduction}

Large data sets, the nourishment of deep learning, are far from disorganized assemblages of numbers:
data has a lot of internal structure.
In fact, data structure is a crucial property
that allows neural networks to learn meaningful features, and ultimately achieve good generalization.
Yet, most existing theoretical results in the statistical mechanics of deep learning overlook 
this inherent characteristic of the training data.
From the classic works of Gardner on classification capacity \cite{Gardner:1987}
to the most recent advances on the structure of the loss landscape \cite{AnnesiLauditi:2023},
inputs are often assumed to be independent and identically distributed random variables.
Statistical learning theory itself, a cornerstone of the theoretical foundations of deep learning,
focuses on worst-case bounds that do not capture adequately the
typical attributes of empirical data 
\cite{Cohn1992,vapnik2013nature,Bottou2015}.

Over the past few years, researchers have started delving into various facets of data structure, 
leading to the definition of several ensembles of structured data 
\cite{Rotondo:2020:PRR, Goldt:2020:PRX, ChungLeeSompolinsky:2018}.
In machine learning, these could be interpreted as simple generative models.
The primary challenge is to construct ensembles that encapsulate interpretable structural elements,
possibly related to established properties of empirical data sets 
\cite{
PetriniCagnetta:2023,
Mazzolini:2018:PRX,
MazzoliniGrilli:2018,
DataComplexityBOOK,GherardiRotondo:2016},
while maintaining a level of simplicity that facilitates rigorous analysis.
(A parallel approach addresses computations for fixed instances of the data
\cite{
beyond_the_infinite_width_limit,
SeroussiNaveh:2023,
QianyiSompolinsky:2021}.)
In the context of disordered systems, data plays the role of disordered interactions \cite{EngelBOOK},
so that the quest for a theory of structured data in machine learning
converges with a more general endeavor in spin glass theory: the exploration of structured disorder 
\cite{Mezard:2023, Gabrie:chapter, Mezard:2017}.

Another necessity in the theory of neural networks, we believe, is the construction of simple models.
We use the word ``model'' in the statistical physics sense:
a model is a simplified representation of a system, designed to capture the essential features
while omitting irrelevant details, and possibly to allow for analytical computations.
A prominent example is the model introduced
by Domany, Meir, and Kinzel (DMK) in \cite{DMK:1986}.
It focuses on the processing of a noisy input, through a fully-connected network of $L$ layers, 
whose parameters are fixed by unsupervised training.
The learning goal is simplified, compared to the loss functions used in practice:
the elements, or patterns, $\xi^0_\mu$ of a training set need to be transformed, at each layer $l$,
into internal representations $\xi^l_\mu$, fixed at the outset.
The set of all these quantities, for $l=0,\ldots,L$, constitutes a specification of the disorder.
The primary question is then how a pattern $\xi^0_\mu$ is processed by
the network when it is corrupted by noise.
Do the noisy internal representations converge, through the layers, to the noiseless $\xi^l_\mu$?

The DMK model was studied, together with a few variants, in 
a series of papers by the original authors and by others
\cite{MeirDomany:1987,DerridaMeir:1988,DKM:1989},
mainly under the denomination of ``layered networks''.
Such models, at the time of their introduction, had the merit of providing a bridge between
the research done by physicists (mainly concerned with associative memory and its dynamical implementation)
and computer scientists \cite{Domany:chapter}.
More recently, Babadi and Sompolinsky \cite{BabadiSompolinsky:2014}, and
Kadmon and Sompolinsky \cite{KadmonSompolinsky:2016} 
analyzed the effect, on such a model, of 
imposing sparsity in the neuronal activations.

In these previous works, the probability measure on the input patterns
$\xi^0_\mu$ and on the internal representations $\xi^l_\mu$ was assumed factorized,
both over patterns and over components.
In this paper we use the DMK model to explore the effects of
data structure in the form of non-null relations between elements of the training set.
The specific model of data structure that we consider is an instance
of two broader classes, manifold learning and simplex learning,
dealing with the classification of extended (non-point) patterns
\cite{
WakhlooSussman:2023,
GherardiEntropy,
RotondoPastore:2020:PRL,
PastoreRotondo:2020:PRE,
Rotondo:2020:PRR,
Borra:2019,
ChungLeeSompolinsky:2018,
ChungLeeSompolinsky:2016}.
The rationale to use such a model is related to its being
(i) the object of existing literature,
(ii) simple enough to be treated analytically,
(iii) a posteriori, complex enough to probe non-trivial
signal-processing features of deep fully-connected networks,
particularly regarding a practically relevant concept:
the resolution of two similar inputs.


\begin{figure}[tb]
\includegraphics[scale=.63]{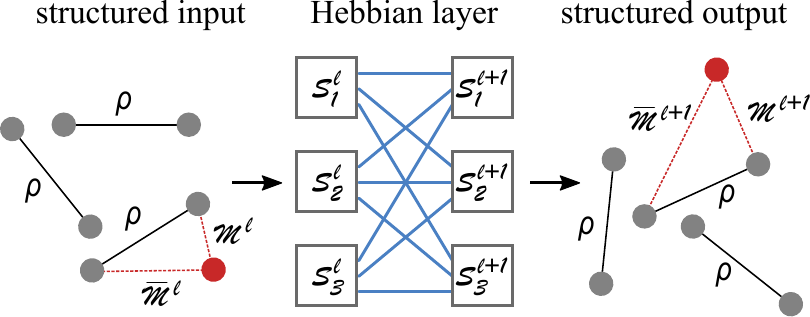}
\caption{
Pictorial description of the model.
Each layer $l$ of a deep fully-connected architecture transforms
(with weights fixed by Hebb's rule) a set of inputs into a set of outputs.
Both these sets are structured as pairs of patterns (grey circles) with prescribed overlaps $\rho$,
Eq.~(\ref{eqn:rho}).
The overlaps $M^l$ and $\bar M^l$ of an input (red circle) with a given pair of patterns
are transformed by the layer into $M^{l+1}$ and $\bar M^{l+1}$.
The equations solving the model express the averages of these order parameters
at layer $l+1$ in terms of those at layer $l$.
}
\label{fig:sketch}
\end{figure}

\section{Solvable model of layer-wise processing of structured data}

In this section, we introduce the model, which describes
the layer-wise processing of structured data
by a network with prescribed internal representations, trained
with Hebb's rule.
Refer to Fig.~\ref{fig:sketch} for a pictorial description of the
salient elements in the model.

\subsection{Formulation of the model}

Consider a network of \(L\) layers, 
each containing \(N\) units with binary values, or activations,
$S_i^l = \pm 1$, with $i=1,\ldots,N$ and $l=1,\ldots,L$.
The activation of unit $i$ in layer $l$ depends on
those of the previous layer
(i.e., the network is feed-forward):
\begin{equation}
\label{eqn:law}
 S_i^l = \sgn \left[ \sum_{j=1}^N W_{ij}^{l-1} \, S_j^{l-1} \right],
\end{equation}
where \(W_{ij}^l\) is a tensor of synaptic weights.
A priori, there are no constraints on $W_{ij}^l$, (i.e., the network is fully connected.)
The activation of each unit is therefore a deterministic function
$S^l_i(\xi)$ of the input $\xi\equiv S^0\in\{-1,1\}^N$.

In supervised learning, both in artificial neural networks
and in models of neural processing of sensory stimuli in the cortex,
the network described by Eq.~(\ref{eqn:law}) is usually followed by
a final readout layer, which would map the $N$-dimensional
representation $S^L$ to a meaningful output.
Here, we avoid modeling this last step, and consider the dynamics
from an unsupervised learning perspective, where the training set $\mathcal T^0$
is a collection of inputs $\{\xi_\mu\}$, with no prescribed labels,
and the learning task is the specification of a fixed number of constraints
\begin{equation}\label{eq:learning_task}
S^l_i (\xi_\mu) = \xi^l_{i\mu}
\end{equation}
at each layer $l$.
We use $\xi^l_{i\mu}$ to denote the data for the $\mu$-th constraint
on the $i$-th unit in the $l$-th layer, 
$\xi^l_\mu$ to denote the vector data for the $\mu$-th constraint
in the $l$-th layer, and occasionally use $l=0$
for the training set elements.
The constraints collectively impose associations between 
input patterns and their internal representations 
(which we refer to as internal patterns) at each layer.
The set of internal patterns 
$\{\xi_1^l, \xi_2^l,\ldots\}$
at layer $l$ is denoted
$\mathcal T^l$.
Altogether, the disorder is specified by
$\mathcal T := \{\mathcal T^0, \mathcal T^1, \ldots, \mathcal T^L\}$.
Note that $\mathcal T^0$, despite its being on an equal footing with
$\mathcal T^{l>0}$, assumes a conceptually distinct role when the model
is viewed within a supervised learning context, because the input data is fixed
whereas in practice the internal patterns are optimized via training.

In order to accomplish the task in Eq.~(\ref{eq:learning_task}),
the weights \(W_{ij}^l\) are fixed by Hebb's rule:
\begin{equation}
\label{eqn:hebb_simple}
W_{ij}^l = \frac{1}{N} \sum_\mu \xi_{i\mu}^{l+1} \xi_{j\mu}^l.
\end{equation}
Although less effective than backpropagation,
this choice of training rule solves each of the constraints in Eq.~(\ref{eq:learning_task})
with probability close to 1, for sufficiently small $P/N$.
The advantage is that it gives an explicit expression for
$W_{ij}^l$ in terms of the disorder.
A more expressive learning rule would be realized by 
solving Eq.~(\ref{eq:learning_task}) via a pseudo-inverse;
the equations, however, can get more complicated,
and the phenomenology is not dissimilar to the Hebbian version
\cite{KadmonSompolinsky:2016, DKM:1989}.

\subsection{Structured disorder}

Our goal is to introduce non-null dependencies between the elements
of the training set and between the internal patterns at each layer.
In particular, we follow a model of data structure that was already
studied in other contexts (see the references cited in the Introduction).
The patterns in $\mathcal T^0$ are organised in pairs, 
$\{\xi_\mu,\bar\xi_\mu\}$, with $\mu=1,\ldots,P$.
The marginal probability distribution
of each component $\xi_{i\mu}$ (and similarly for $\bar\xi_{i\mu}$) is uniform on $\{-1,1\}$.
Patterns belonging to different pairs are independent, 
while patterns within a pair have, on average, a fixed overlap \(\rho\),
independently of $\mu$:
\begin{equation}\label{eqn:rho}
\frac{1}{N} \sum_{i=1}^{N} \left<\xi_{i\mu} \bar\xi_{i\mu}\right> = \rho.
\end{equation}
The scalar $\rho$ parameterizes the disorder,
and will be called structure parameter.
Averages $\left<\cdot\right>$ are over the disorder, i.e.,
integrals over the probability measure of 
all patterns in $\mathcal T$.

This model of data structure fixes the measure of $\mathcal T^0$ only.
To completely specify the disorder measure in the DMK model,
we use the same prescription for each set of internal representations
$\mathcal T^l$ (with $l=0,\ldots,L$)
where any two sets $\mathcal T^{l_1}, \mathcal T^{l_2}$
with $l_1\neq l_2$ are independent.
Beyond analytic tractability, the rationale for such a choice is
the assumption that correlations, here in the specific form of a fixed overlap,
are an important feature of the input data that should be
propagated through the network.

With these choices, Hebb's rule, Eq.~(\ref{eqn:hebb_simple}), reads
\begin{equation}
\label{eqn:hebb}
W_{ij}^l = \frac{1}{N} \sum_{\mu=1}^P \left(\xi_{i\mu}^{l+1} \xi_{j\mu}^l + \bar\xi^{l+1}_{i\mu} \bar\xi_{j\mu}^{l} \right).
\end{equation}

\subsection{Choice of order parameters}

To follow the propagation of the input state $S^0$ through the network, 
we consider the overlap $M_\mu^l$ between the internal state $S^l$ at layer $l$
and the internal pattern $\xi^l_\mu$
(and similarly for $\bar \xi^l_\mu$):
\begin{equation}\label{eq:def_M}
\begin{split}
M_\mu^l &= \frac{1}{N}\sum_{i=1}^N S_i^l\,\xi_{i\mu}^l \\
\bar M_\mu^l &= \frac{1}{N}\sum_{i=1}^N S_i^l \, \bar\xi_{i\mu}^l.
\end{split}
\end{equation}
We assume that the input $S^0$ has a finite overlap with exactly one pair in $\mathcal T^0$
(fixed to $\mu=1$) and is independent of the patterns $\xi^0_\mu,\bar\xi^0_\mu$ with $\mu>1$.
Then
\begin{equation}\label{eqn:inital_M}
M_\mu^0 ,\, \bar M_\mu^0 =
\begin{cases*}
O(1) & if $\mu=1$ \\
O(1/\sqrt{N}) & otherwise.
\end{cases*}
\end{equation}
Solving the model, with respect to these order parameters, means
calculating the average overlaps
\begin{equation}
\begin{split}
m^l &= \left< M_1^l \right>\\
\bar m^l &= \left< \bar M_1^l \right>.
\end{split}
\end{equation}
We will achieve this by writing recurrence relations, i.e.,
expressions of $m^{l+1}$ and $\bar m^{l+1}$ in terms of
the order parameters at layer $l$.
A complete set of order parameters, such that the final equations are closed,
is described in the next section.
Since only layers $l$ and $l+1$ will appear in the calculations,
we will simplify the notation and suppress the index for the layer \(l\), while
all variables associated to layer \(l+1\) will be primed;
for instance, $M_\mu \equiv M^l_\mu$ and $M'_\mu \equiv M^{l+1}_\mu$.

\subsection{Mean-field solution in the thermodynamic limit}\label{sec:solution}

Our goal is to find a set of order parameters 
$\bm Q \equiv \{Q_i\}$, $i=1,\ldots, N_Q$,
such that $Q_1 \equiv m$ and $Q_2 \equiv \bar m$, satisfying closed recurrence relations
\begin{equation}
Q'_i = F_i\left(Q_1, Q_2, \ldots, Q_{N_Q}\right) 
\end{equation}
in the thermodynamic limit
\begin{equation}\label{eq:thermodynamic_limit}
N\to\infty, \quad P\to\infty, \quad \alpha:=\frac{P}{N} \;\; \mathrm{finite}.
\end{equation}
This limit is known as the proportional limit, or the finite-width regime,
in the machine learning literature
\cite{beyond_the_infinite_width_limit, ZavatoneVethCanatar:2022}.
The computation leading to a feasible choice of $\bm Q$ and
the corresponding functions $F_i$ is sketched here; 
the details are reported in the Appendix.

An attempt to express $m'$ (or equivalently $\bar m'$) in terms of $m$ and $\bar m$
lands on
\begin{equation}\label{eq:sketch_m}
\begin{split}
m' &= \left< \sgn \left[m+\theta \bar m + x\right] \right>\\
x &= \sum_{\mu\neq 1}
\left(\eta_\mu M_\mu+\bar\eta_\mu\bar M_\mu\right),
\end{split}
\end{equation}
where $\theta$, $\{\eta_\mu\}$, $\{\bar\eta_\mu\}$ are $2P+1$
random variables in $\{-1,+1\}$,
all uncorrelated with each other except pairs $\{\eta_\mu,\bar\eta_\mu\}$,
and uncorrelated with $M_\mu$ and $\bar M_\mu$,
with averages $\left<\theta\right>=\rho$ and 
$\left<\eta_\mu\right>=\left<\bar\eta_\mu\right>=0$.
To proceed, we make two assumptions:
(1) we assume, self-consistently, that Eq.~(\ref{eqn:inital_M})
holds more in general for $M^l_\mu$ and $\bar M^l_\mu$, at each layer $l$;
(2) we assume that $x$ is a normal random variable,
$x\sim\mathcal N(0,\Delta)$.
The variance of $x$,
\begin{equation}
\Delta= \underbrace{\sum_{\mu \neq 1} \left< M_{\mu}^2 + \bar M_{\mu}^2  \right>}_{\Deltad}
+ \underbrace{2\rho \sum_{\mu \neq 1} \left<M_{\mu} \bar M_{\mu}\right>}_{\Deltac},
\end{equation}
is the sum of a diagonal term $\Deltad$ and an off-diagonal (cross) term $\Deltac$,
which are then natural candidates for the missing order parameters:
$Q_3\equiv\Deltad$, $Q_4\equiv\Deltac$.
It is straightforward to compute the average in Eq.~(\ref{eq:sketch_m})
as an integral over $x$ and a sum over $\theta=\pm 1$.
The computation of $\Deltad'$ and $\Deltac'$
is slightly more involved, as it includes a larger set
of random variables, but it does not present any additional difficulties.
Finally one gets the following recurrence equations
for the 4 order parameters $\{m,\bar m,\Deltad,\Deltac\}$:

\begin{equation}
\label{eqn:recursive_m}
m' = \frac{1+\rho}{2}\erf \left( \frac{m+\bar m}{\sqrt{2\Delta}} \right) + \frac{1-\rho}{2}\erf\left( \frac{m-\bar m}{\sqrt{2\Delta}} \right)
\end{equation}
\begin{equation}
\label{eqn:recursive_barm}
\bar m' = \frac{1+\rho}{2}\erf\left( \frac{\bar m+m}{\sqrt{2\Delta}} \right) + \frac{1-\rho}{2}\erf \left( \frac{\bar m-m}{\sqrt{2\Delta}} \right)
\end{equation}
\begin{equation}
\label{eqn:recursive_deltad}
 \Deltad'
 = 2\alpha + \frac{2}{\pi} \frac{\Omega_\rho(m,\bar m,\Delta)}{\Delta} 
 \left[ (1+\rho^2)\Deltad+2\Deltac
 \right]
\end{equation}
\begin{equation}
\label{eqn:recursive_deltac}
\Deltac'
 = 2\alpha\rho^2 + \frac{2}{\pi} \frac{\Omega_\rho(m,\bar m,\Delta)}{\Delta} 
 \left[ 2\rho^2\Deltad+(1+\rho^2)\Deltac
 \right]
\end{equation}
where
\begin{equation}\label{eq:omega}
\begin{split}
\Omega_\rho(m,\bar m,\Delta) :=
\frac{(1+\rho)^2}{4} e^{-\frac{\left(m+\bar m\right)^2}{\Delta}}
&+\frac{(1-\rho)^2}{4} e^{-\frac{\left(m-\bar m\right)^2}{\Delta}}\\
&+\frac{1-\rho^2}{2} e^{-\frac{m^2+\bar m^2}{\Delta}}.
\end{split}
\end{equation}
(In the formulas above, $\mathrm{erf}$ is the standard error function.)
By using the shorthand $\bm Q=(m,\bar m,\Deltad,\Deltac)$, the recurrence equations can be written
compactly as
\begin{equation}\label{eq:recurrence_compact}
\bm Q'= F(\bm Q),
\end{equation}
where $F$ is given by Eqs.~(\ref{eqn:recursive_m}--\ref{eqn:recursive_deltac}).

The initial conditions for $\Deltad$ and $\Deltac$ can be obtained from
Eq.~(\ref{eq:def_M}) and the assumption that $S^0$ is independent from $\xi^0_\mu,\bar\xi^0_\mu$ with $\mu>1$.
They are fixed by the parameters $\alpha$ and $\rho$:
\begin{equation}\label{eq:delta_initial}
\begin{split}
\Deltad^{l=0}&=2\alpha\\
\Deltac^{l=0}&=2\alpha\rho^2,
\end{split}
\end{equation}
while $m^{l=0}$ and $\bar m^{l=0}$ can be varied
(see Sec.~\ref{sec:basins}).

\begin{figure*}[t]
\includegraphics[scale=.64]{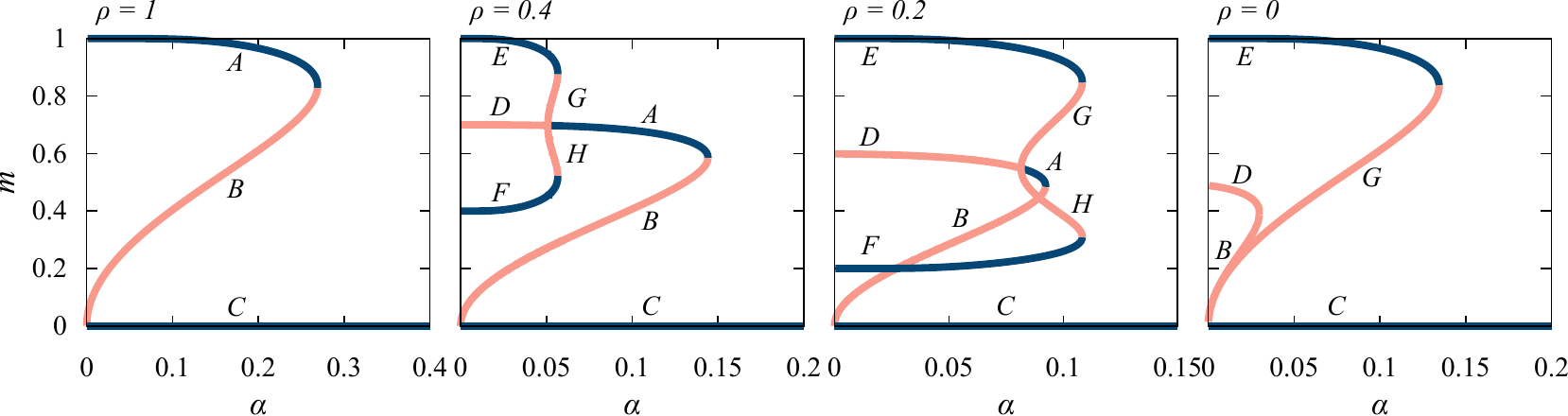}
\caption{
Bifurcation diagrams at four different values of the structure parameter $\rho$,
showing the overlap $m$ at all fixed points (y axis), as a function of the load $\alpha$ (x axis).
Stable branches are in blue and unstable branches in pink.
Letters are the names given in the text to the different branches.
For $\rho=1$ the diagram is identical to the one in the original, unstructured, DMK model,
with two stable and one unstable branches.
When data is structured, $\rho<1$, the diagram is topologically different,
and more branches appear.
Note that these are 2-dimensional projections of 4-dimensional bifurcation diagrams.
For instance, branches $F$ and $B$ never intersect in the 4-dimensional
space, even though they do so in the plot for $\rho=0.2$.
}
\label{fig:bifurcations}
\end{figure*}

\subsection{Special limits}\label{sec:limits}

The model becomes equivalent to the one with unstructured disorder
in the two distinct limits $\rho\to 0,1$.

First, when $\rho=1$, the right-hand sides of
Eqs.~(\ref{eqn:recursive_m}) and (\ref{eqn:recursive_barm})
both become equal to $\erf (m+\bar m)/\sqrt{2(\Deltad+\Deltad)}$.
Since the initial conditions for $m$ and $\bar m$ are the same,
then $m^l=\bar m^l$ for all $l$.
The same happens with Eqs.~(\ref{eqn:recursive_deltad}) and (\ref{eqn:recursive_deltac}),
which become
$\Deltad'=\Deltac'=4\alpha+8/\pi \exp[-4(m+\bar m)^2/(\Deltad+\Deltac)]$.
Again, the initial conditions (\ref{eq:delta_initial}) are equal for $\rho=1$, therefore
$\Deltad^l=\Deltac^l$.
Finally, by recognizing $\Delta\equiv (\Deltad+\Deltac)/4$, one obtains
the unstructured recurrence relations
\begin{equation}\label{eqn:recursion_unstructured}
\begin{split}
m' &= \erf \frac{m}{\sqrt{2\Delta}}\\
\Delta' &= \alpha + \frac{2}{\pi} e^{-\frac{m^2}{\Delta}}
\end{split}
\end{equation}
with initial conditions
$m^{l=0}=M_1^1$, $\Delta^{l=0}=\alpha$.

Second, when $\rho=0$ and the initial condition is $\bar m^{l=0}=0$,
Eq.~(\ref{eqn:recursive_barm}) gives $\bar m'=0$, whence
$\bar m^l=0$.
Equation (\ref{eqn:recursive_deltac}) reduces to $\Deltac'\propto\Deltac$,
thus $\Deltac^l=0$, since $\Deltac^0=0$.
Specializing Eq.~(\ref{eqn:recursive_deltad}) to these values,
with $\Omega_{\rho=0}(m,0,\Delta,0) = \exp{(-m^2/\Delta)}$,
and recognizing $\Delta \equiv \Deltad$,
one then obtains Eqs.~(\ref{eqn:recursion_unstructured}) with the substitution $\alpha \leftarrow 2\alpha$.
The initial condition in this case is
$m^{l=0}=M_1^1$, $\Delta^{l=0}=2\alpha$.
The doubling of $\alpha$ reflects the doubled number of points.

Ultimately, it is interesting to view this model as an interpolation between
two models with unstructured disorder, one where pairs of points are indistinguishable ($\rho=1$)
and one where pairs of points are orthogonal on average ($\rho=0$).

\subsection{Symmetries}

The recurrence equation (\ref{eq:recurrence_compact}) governing the discrete-time dynamics in the model
satisfies two notable symmetries, namely swapping the overlaps $m$ and $\bar m$,
and taking their negative values.
More precisely, the two functions
$\sigma$ and $\tau$, defined as
\begin{equation}
\begin{split}
\sigma(\bm Q) &:=
\begin{pmatrix}
0 & 1 &0 &0 \\
1 & 0 & 0 & 0 \\
0 & 0 & 1 & 0 \\
0 & 0 & 0 & 1
\end{pmatrix}
\times \bm Q\\
\tau(\bm Q) &:=
\begin{pmatrix}
-1 & 0 &0 &0 \\
0 & -1 & 0 & 0 \\
0 & 0 & 1 & 0 \\
0 & 0 & 0 & 1
\end{pmatrix}
\times \bm Q,
\end{split}
\end{equation}
satisfy $[\sigma,F]=[\tau,F]=0$, or in other terms
\begin{equation}
\begin{split}
\sigma(\bm Q')&=F(\sigma(\bm Q)) \\
\tau(\bm Q')&=F(\tau(\bm Q)).
\end{split}
\end{equation}
We call $\sigma$ and $\tau$ exchange and reflection symmetry respectively.

\section{Results}

\subsection{Bifurcation diagram}\label{sec:bifurcations}

The order parameters at the output layer of an $L$-layer network are obtained
by iterating Eqs.~(\ref{eqn:recursive_m}--\ref{eqn:recursive_deltac}) $L$ times.
Here we focus on the infinite-depth limit, where $L\to\infty$ 
\cite{PeluchettiFavaro:2020,Hanin:2021}.
In this limit, the values attainable, at fixed $\alpha$, by the order parameters are
reduced to a finite number, since the dynamics converges to stable fixed points.
It is therefore interesting to consider the fixed points of the dynamics, i.e.,
sets of order parameter values such that 
$(m',\bar m', \Deltad', \Deltac')=(m,\bar m, \Deltad, \Deltac)$.
We will plot $m$ (equivalently, $\bar m$) of these fixed points as functions of the parameter $\alpha$,
for fixed $\rho$, and refer to these plots as \textit{bifurcation diagrams},
since, as we checked numerically,
the dynamics considered here has no periodic orbits and no chaotic attractors.

Note that, thanks to the reflection symmetry $\tau$, we can restrict the analysis to
non-negative values of $m$. We will plot the bifurcation diagrams in this range,
and it is understood that the corresponding diagrams for $m\in[-1,0]$
are the mirror images of those.

The bifurcation diagram for the unstructured case was described
in Ref.~\cite{MeirDomany:1987}.
We reproduce it in Fig.~\ref{fig:bifurcations} by taking $\rho=1$ in our equations.
For all $\alpha$, there is a stable fixed point with $m=0$ (branch $C$).
For $\alpha<\alpha_\mathrm{c} \approx 0.269$ two more branches appear,
a stable one (branch $A$) that converges to $m=1$ for $\alpha=0$,
and an unstable one (branch $B$) that merges the two stable branches.
The critical point $\alpha_\mathrm{c}$ is a fold bifurcation,
where a stable fixed point and an unstable one merge to a saddle point
\cite{Kuznetsov:BOOK}.
Since this critical point is present for all $\rho$, we will use the
notation $\alpha_\mathrm{c}(\rho)$; so $\alpha_\mathrm{c}=\alpha_\mathrm{c}(1)$.

Now, in the structured case, when $\rho\neq0,1$, is the bifurcation
diagram topologically equivalent to the unstructured one?
Interestingly, the answer is no. 
This fact can be appreciated analytically;
we postpone the details to Sec.~\ref{sec:symmetry_breaking}, where we consider the limit $\alpha\to 0$.
Here we compute the diagram numerically.
The full bifurcation diagrams for $\rho=1,0.4,0.2,0$,
obtained by solving the fixed-point equations 
with Newton's method, are shown in Fig.~\ref{fig:bifurcations}.
To evaluate the stability of the branches, we computed
the largest eigenvalue $\lambda_\mathrm{max}$ of the Jacobian matrix
$J$ with components
\begin{equation}
\begin{split}
&J_{ij}=\frac{\partial x'_i}{\partial x_j}, \quad i,j=1,\ldots,4\\
&x=(m,\bar m,\Deltad,\Deltac); \quad
x'=(m',\bar m',\Deltad',\Deltac'),
\end{split}
\end{equation}
evaluated at the fixed points.
If $\lambda_\mathrm{max}<1$ (respectively $>1$), the fixed point is stable
(respectively unstable). We do not consider the marginal case $\lambda_\mathrm{max}=1$.

\begin{figure}[t]
\includegraphics[scale=.57]{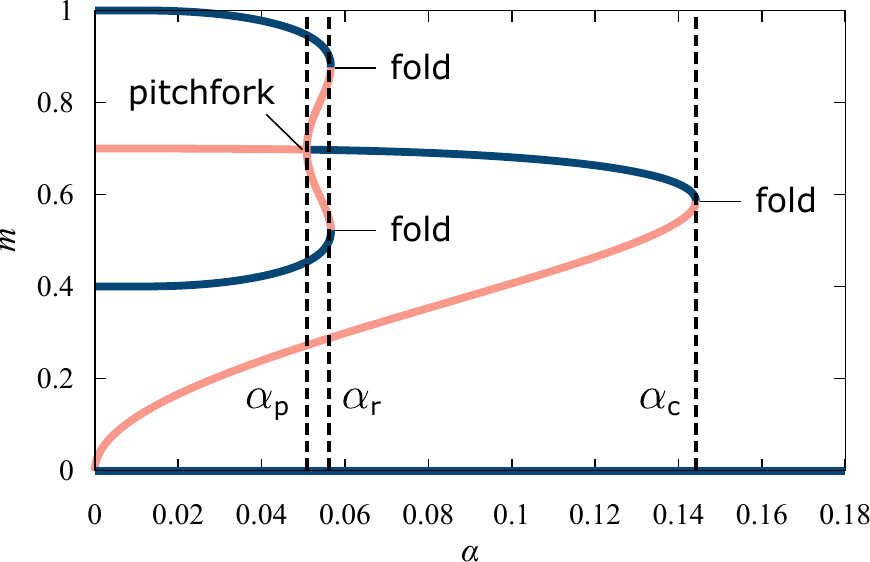}
\caption{
Bifurcations and critical loads (here for the case $\rho=0.4$).
From small to large values of the load $\alpha$ (x axis),
the order parameter $m$ at the fixed points (y axis)
undergoes a pitchfork bifurcation at $\alpha_\mathrm{p}$,
two fold bifurcations at $\alpha_\mathrm{r}$,
and a single fold bifurcation at $\alpha_\mathrm{c}$.
}
\label{fig:bifurcations_onerho}
\end{figure}

As soon as $\rho<1$, another bifurcation appears, at $\alpha=\alpha_\mathrm{p}(\rho)$;
for instance, Fig.~\ref{fig:bifurcations_onerho} shows that $\alpha_\mathrm{p}(0.4)\approx 0.05$.
At this special point, the local structure of the system transitions from a single unstable fixed point
(for $\alpha<\alpha_\mathrm{p}$) to a pair of unstable points plus a stable one
(for $\alpha>\alpha_\mathrm{p}$).
The point where this happens is commonly referred to as a (subcritical) pitchfork bifurcation, 
hence the subscript of $\alpha_\mathrm{p}$.
Such a bifurcation usually occurs in systems with symmetries;
in this case, the symmetry involved is $\sigma$, as discussed in the next section.
Connected to this bifurcation, there are 3 branches with no topologically
equivalent branches in the unstructured case;
these are branches $D$, $G$, and $H$.
In turn, branches $G$ and $H$ give rise to two more fold bifurcations,
where they merge with stable branches $E$ and $F$.
Importantly, the two fold bifurcations happen at the same value of $\alpha$,
which we call $\alpha_\mathrm{r}(\rho)$ for reasons that will be clear in the following.
The types of the bifurcations and their locations in terms of $\alpha$
are summarized in Fig.~\ref{fig:bifurcations_onerho}.

The value of $\alpha_\mathrm{p}(\rho)$ gets larger for smaller $\rho$:
branch $A$ shrinks and branches $E$ and $F$ expand;
see the rightmost panels in Fig.~\ref{fig:bifurcations}.
Note that this plot shows a 2-dimensional projection 
of a 4-dimensional bifurcation diagram,
therefore branches that intersect in the figure, such as $F$ and $B$, need not
do so in the full space.
In fact, the topology of the bifurcation diagram remains the same
for all $\rho\in(\rho_\mathrm{c},1)$, with $\rho_\mathrm{c}\approx 0.12$.

As $\rho$ decreases further, the bifurcation at $G\cap H$ moves rightward,
up to $\rho=\rho_\mathrm{c}$ when branch $A$ disappears.
As a consequence, when $\rho<\rho_\mathrm{c}$, the bifurcation
at $G\cap H$ now involves only unstable fixed points,
and the same holds for the bifurcation at $D\cap B$, which takes the place
of $A\cap B$.
When decreasing $\rho$ even further, the point $G\cap H$ moves leftward
along branch $B$, and the part of the diagram that was made of
branches $D\cup A\cup B$ for $\rho>\rho_\mathrm{c}$,
which is now $D\cup B$, shrinks towards smaller values of $\alpha$.
Finally, approaching $\rho= 0$, where the model 
reproduces the unstructured one (as explained in Sec.~\ref{sec:limits}), branches $H$ and $F$
get squashed onto the x axis, and branches $E$ and $G$
reproduce the unstructured branches $A$ and $B$.

\subsection{Symmetry-breaking fixed points and the resolution transition}\label{sec:symmetry_breaking}

Let us consider here the limit $\alpha\to 0$,
where the nature of the stable branches $E$ and $F$ in the bifurcation diagram
can be studied analytically.
In this limit,
the erf functions in Eqs.~(\ref{eqn:recursive_m})
and (\ref{eqn:recursive_barm}) saturate, so that
$(m,\bar m) \approx (1,\rho)$ is a possible solution of the fixed point equation.
The initial condition sets $\Deltad^{l=0}+\Deltac^{l=0}=2\alpha(1+\rho^2)$,
therefore the terms containing $\Omega$ are exponentially suppressed in this limit.
These observations suggest the following Ansatz:
\begin{equation}\label{eq:ansatz}
\begin{split}
&m=1-\epsilon,\quad \bar m=\rho+\epsilon,\\
&\Delta_\mathrm{d}=2\alpha, \quad \Delta_\mathrm{c}=2\alpha\rho^2,
\end{split}
\end{equation}
with $\epsilon>0$.
Substituting (\ref{eq:ansatz}) in Eqs.~(\ref{eqn:recursive_m}) and (\ref{eqn:recursive_barm}),
and using the approximation $\erf(x)\approx 1 - 1/\sqrt{\pi} e^{-x^2}/x$,
valid for large $x$, gives expressions for $m'$ and $\bar m'$ containing
terms proportional to $\exp(-x_+^2)$ and $\exp(-x_-^2)$, with arguments
$x_+:=(1+\rho)/\sqrt{2\alpha(1+\rho^2)}$ and $x_-:=(1-\rho-2\epsilon)/\sqrt{2\alpha(1+\rho^2)}$.
Supposing $\rho>0$, one has that $x_+>x_-$.
Neglecting the subleading exponentials, one then obtains
$(m',\bar m')=(1-h(\epsilon; \alpha, \rho), \rho+h(\epsilon; \alpha, \rho))$, with
\begin{equation}\label{eq:function_h}
h(\epsilon; \alpha, \rho) = \frac{1-\rho}{2\sqrt{\pi}} \frac{1}{x_-}\exp(-x_-^2).
\end{equation}
Therefore, the point $(m,\bar m,\Deltad,\Deltac)$ is a fixed point
when $\epsilon=h(\epsilon; \alpha, \rho)$.
At fixed $\alpha$ and $\rho$, $h(\epsilon; \alpha, \rho)$ is a convex positive function of $\epsilon$,
which diverges at $\epsilon=(1-\rho)/2$, is punctually monotonically increasing with $\alpha$,
and is such that $h(0; \alpha\to 0, \rho)\to 0$.
Thus, for small enough $\alpha$, the equation $\epsilon=h(\epsilon; \alpha, \rho)$
will have two solutions, $\epsilon_\pm$, such that
$\epsilon_+\to (1-\rho)/2$ and $\epsilon_-\to 0$ for $\alpha\to 0$.

The solution $\epsilon_+$ gives a fixed point that is, at least for $\alpha=0$,
symmetric: $\left.m\right|_{\epsilon_+}=\left.\bar m\right|_{\epsilon_+}=(1+\rho)/2$.
This corresponds to an infinite-depth limit where the output of the network
does not distinguish between the two partners in the same pair.
The numerical analysis of Fig.~\ref{fig:bifurcations}
shows that this is actually a whole branch
extending to finite values of $\alpha$ (branch $D$ in the bifurcation diagrams).
The solution $\epsilon_-$, instead, gives a fixed point where 
$m\approx 1$ and $\bar m\approx \rho$.
This asymmetric fixed point is
present for positive values of $\alpha$, thus giving rise
to a new branch in the bifurcation diagram.
Because of the exchange symmetry $\sigma$,
this implies the existence of two symmetric branches
(branches $E$ and $F$ in Fig.~\ref{fig:bifurcations}).
These branches are attractors of a dynamics in which 
differential information about the two correlated patterns in the pair
is not lost, and the state $S^L$
stays close to the $L$-layer representation of the input pattern, within the pair 
$\{\xi^0_1,\bar\xi^0_1\}$,
that was closer to the input $S^0$
(more details are given in the next section).
We can interpret this regime by stating that the network
can resolve patterns having overlap $\rho$.
Hence, we call the critical value $\alpha_\mathrm{r}(\rho)$,
below which this is possible, the ``resolution'' transition.
For $\alpha_\mathrm{r}(\rho)<\alpha<\alpha_\mathrm{c}(\rho)$, the network's output
retains a positive overlap with both patterns in the pair,
but it cannot resolve which one was closer to the input.

The analysis based on the ansatz (\ref{eq:ansatz}) also suggests that the branches $E$ and $F$
should disappear for sufficiently large $\alpha$, when the graph of the function $h$,
Eq.~(\ref{eq:function_h}),
lies completely above the bisector
between $\epsilon=0$ and the singularity $\epsilon=(1-\rho)/2$.
This is indeed the case for the diagrams in Fig.~\ref{fig:bifurcations}.
The approximation, however, is too crude to yield a quantitatively
useful estimate of $\alpha_\mathrm{r}(\rho)$.

\begin{figure}[tb]
\includegraphics[scale=.65]{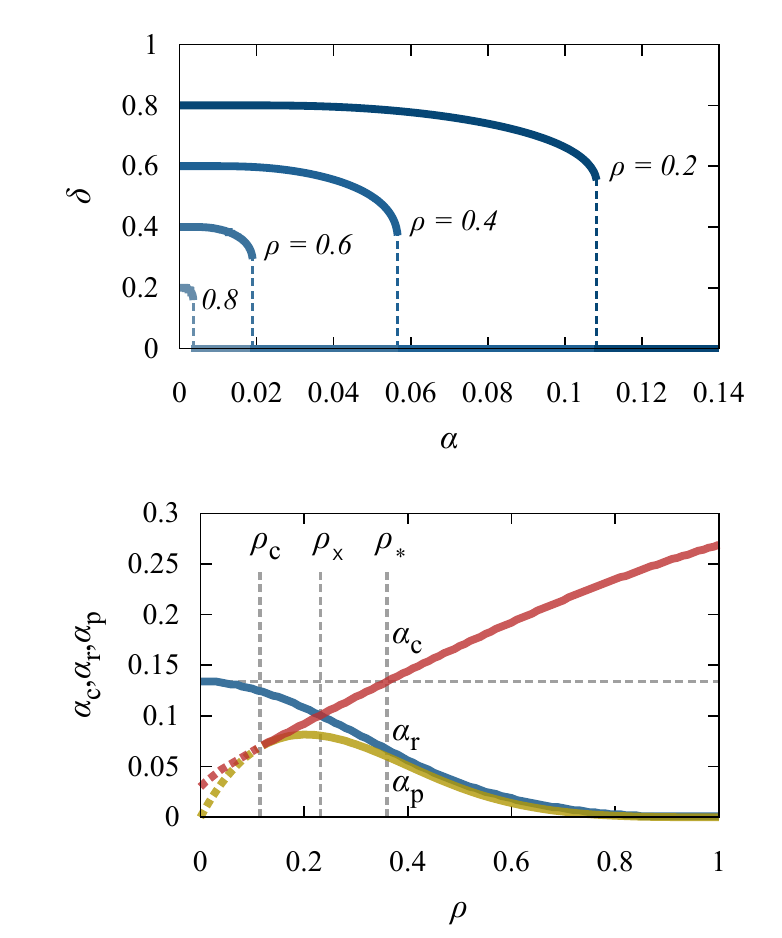}
\caption{
Top panel: the order parameter $\delta(\alpha,\rho)$ (y axis) of the symmetry-breaking transition 
that occurs at $\alpha_\mathrm{r}(\rho)$.
The transition is discontinuous, and
shifts to larger loads $\alpha$ (x axis) as $\rho$ is decreased.
Bottom panel: the three critical points $\alpha_\mathrm{c}$, $\alpha_\mathrm{r}$, and $\alpha_\mathrm{p}$
(y axis), as functions of the structure parameter $\rho$ (x axis).
Below $\rho_\mathrm{c}$, the graphs of $\alpha_\mathrm{d}$ and $\alpha_\mathrm{p}$
are dashed, indicating that the two bifurcations involve only unstable points.
The other special values of $\rho$ are $\rho_\times$, where $\alpha_\mathrm{r}$ and $\alpha_\mathrm{c}$
cross, and $\rho_*$, where $\alpha_\mathrm{c}$ is equal to $\alpha_\mathrm{r}(\rho=0)$.
}
\label{fig:criticalpoints}
\end{figure}

%
Below the resolution transition $\alpha_\mathrm{r}(\rho)$, the symmetry $\sigma$ 
can be broken.
Let $\mathcal S(\alpha, \rho)$ be the set of stable fixed points at given $\alpha, \rho$.
A compact order parameter for this transition is
\begin{equation}
\begin{split}
&\delta(\alpha,\rho) := \sup_{(m,\bar m,\Deltad, \Deltac)\in\mathcal S(\alpha,\rho)} 
\hat\delta(m,\bar m,\Deltad, \Deltac) \\
&\hat\delta(m,\bar m,\Deltad, \Deltac) := \left|m-\bar m\right|.
\end{split}
\end{equation}
Above the transition, the solutions of the equations are symmetric,
meaning that $\hat\delta(x)=\hat\delta(\sigma(x))$,
in terms of the vector $x=(m,\bar m,\Deltad, \Deltac)$.
Thus, in this regime one has that $\delta(\alpha, \rho)=0$.
Below the transition, there is always a stable branch where $m$ and $\bar m$ can take different values,
and $\delta(\alpha,\rho)>0$.
Figure \ref{fig:criticalpoints} shows
that, in terms of this order parameter, the transition can be characterized
as a discontinuous one. The order parameter is such that, for all $\rho$,
$\delta(\alpha,\rho)=0$ when $\alpha>\alpha_\mathrm{r}(\rho)$, and
\begin{equation}
\lim_{\alpha\to\alpha_\mathrm{r}(\rho)^-} \delta(\alpha,\rho) > 0.
\end{equation}

\subsection{Crossing of critical loads}

The phenomenology described above implies a non trivial flow
of the three critical points 
$\{\alpha_\mathrm{c}, \alpha_\mathrm{d}, \alpha_\mathrm{p}\}$
as functions of the structure parameter $\rho$.
The three functions $\alpha_\mathrm{c,d,p}(\rho)$ are plotted in Fig.~\ref{fig:criticalpoints}.

The critical $\alpha_\mathrm{c}$ monotonically increases with $\rho$,
from $\alpha_\mathrm{c}(0) = 0.030(1)$
to $\alpha_\mathrm{c}(1) = 0.269(1)$, which is the critical load of the unstructured case.
The resolution transition, $\alpha_\mathrm{r}$, instead,
decreases from $\alpha_\mathrm{r}(0)=0.134(1)$ to $\alpha_\mathrm{r}(1)=0$.
As shown in Sec.~\ref{sec:limits}, the bifurcation diagram
for $\rho=0$ can be identified with the one for $\rho=1$ by the
mapping $\bar m\mapsto 0$, $\alpha \mapsto 2\alpha$.
Remember also from Sec.~\ref{sec:bifurcations}, that it is one of the asymmetric branches related to the
resolution transition that reproduces, for $\rho\to 0$, the unstructured branch
that extends up to $\alpha_\mathrm{c}(1)$.
Coherently with this equivalence, the numerical values of the critical points satisfy
the relation
\begin{equation}
\alpha_\mathrm{c}(1)=2\alpha_\mathrm{r}(0).
\end{equation}

There are two special values of the structure parameter:
$\rho_\mathrm{c}$ and $\rho_\times$.
They can be defined as follows:
\begin{equation}
\begin{split}
\alpha_\mathrm{c}(\rho_\mathrm{c}) &= \alpha_\mathrm{p}(\rho_\mathrm{c})\\
\alpha_\mathrm{c}(\rho_\times) &= \alpha_\mathrm{r}(\rho_\times)
\end{split}
\end{equation}
At the point $\rho_\times$, the critical loads $\alpha_\mathrm{c}$
and $\alpha_\mathrm{r}$ cross;
at $\rho_\mathrm{c}$, the critical loads $\alpha_\mathrm{c}$
and $\alpha_\mathrm{p}$ touch without crossing.
The special values are $\rho_\mathrm{c}=0.12$ and $\rho_\times=0.23$.

We have defined the critical loads in terms of the locations of the bifurcations.
Alternatively, one may define them in terms of the symmetry of the
stable fixed points, which are those generically reached by the dynamics
(see also the next section for a discussion of the dependence of these
from the initial condition).
Above $\alpha_\mathrm{c}$, the only symmetric stable fixed point has $m=\bar m=0$.
This is true for all $\rho$, both greater and less than $\rho_\times$, i.e., regardless
of whether $\alpha_\mathrm{c}\lessgtr\alpha_\mathrm{r}$.
Above $\alpha_\mathrm{r}$, there are no stable fixed points with broken symmetry,
and below $\alpha_\mathrm{p}$, there are no non-zero symmetric stable fixed points
(again independently of $\rho$).

Let $\mathcal S_\mathrm{r}(\alpha, \rho)$ and $\mathcal S_\mathrm{\bar r}(\alpha, \rho)$ 
be the two subsets of $\mathcal S(\alpha, \rho)$
such that $\hat\delta>0$ and $\hat\delta=0$ respectively.
In other words, $\mathcal S_\mathrm{r}(\alpha, \rho)$ is the subset of stable fixed points
that are not symmetric with respect to $\sigma$, and
$\mathcal S_\mathrm{\bar r}(\alpha, \rho)$ is the subset of stable fixed points
that are symmetric.
The subscript stands for ``resolution'', since, as explained above, the symmetric fixed points
correspond to a situation where the network cannot resolve the two patterns in the pair.
Define also the set $\mathcal S_0(\alpha, \rho) \subseteq \mathcal S_\mathrm{\bar r}(\alpha, \rho)$
as the set of stable fixed points such that $m=\bar m=0$.
Then the critical loads can be defined as follows:
\begin{equation}\label{eq:def_alphas}
\begin{split}
\alpha_\mathrm{c}(\rho)&:=\inf \left\{ \alpha_* \,|\, \mathcal S_\mathrm{\bar r}(\alpha,\rho)
=\mathcal S_0(\alpha, \rho) \:\mathrm{for}\: \alpha>\alpha_*\right\} \\
\alpha_\mathrm{r}(\rho)&:=\inf \left\{ \alpha_* \,|\, \mathcal S_\mathrm{r}(\alpha,\rho)=\emptyset
\:\mathrm{for}\: \alpha>\alpha_*\right\} \\
\alpha_\mathrm{p}(\rho)&:=\inf \left\{ \alpha \,|\, \mathcal S_\mathrm{\bar r}(\alpha,\rho)
\neq \mathcal S_0(\alpha, \rho) \right\} \\
\end{split}
\end{equation}
The definitions for $\alpha_\mathrm{c}$ and $\alpha_\mathrm{p}$ 
coincide with those given in terms of bifurcations
only for $\rho>\rho_\mathrm{c}$. Below $\rho_\mathrm{c}$,
the relevant bifurcations involve only unstable branches, and the definitions disagree.
In particular, $\alpha_\mathrm{c}$ becomes $0$ and $\alpha_\mathrm{p}$ is undefined. 
(In Fig.~\ref{fig:criticalpoints}, the graphs of these two critical loads 
are dashed in the region $\rho<\rho_\mathrm{c}$.)

\begin{figure*}[tb]
\includegraphics[scale=.41]{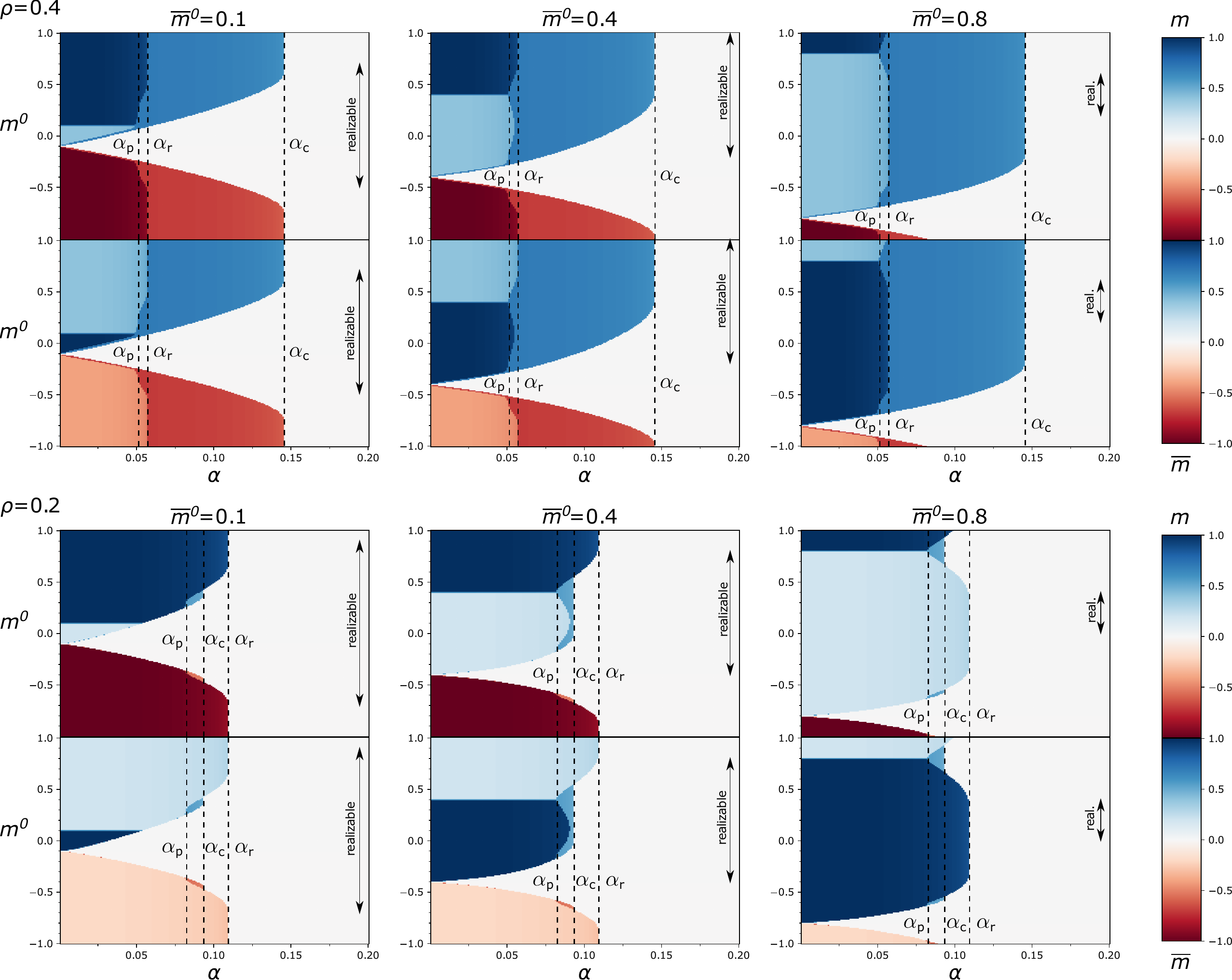}
\caption{
Asymptotic fixed points reached by the dynamics as functions
of the initial condition and of the load $\alpha$ (x axes), 
for $\rho=0.4$ (top panel) and $\rho=0.2$ (bottom panel).
The initial condition is fixed by $m^0$ (y axes) and $\bar m^0$, which is
equal to $0.1$, $0.4$, and $0.8$ in the leftmost, center, and rightmost column respectively.
Colors correspond to the value of $m$ or $\bar m$ at the fixed point reached asymptotically,
here approximated by the 1000-th iterate of the map (\ref{eq:recurrence_compact}).
In each of the two panels, the top row of plots shows $m$ and the bottom one $\bar m$.
Vertical arrows indicate the region of geometrically allowable values of $m^0$, 
i.e., those satisfying Eq.~(\ref{eq:constraints}),
for each of the conditions.
}
\label{fig:basins}
\end{figure*}

\subsection{Phase diagram of attraction basins}\label{sec:basins}

The previous discussion focused on the bifurcation diagram,
which is independent of the initial conditions.
Of course, there are relevant aspects of the dynamics that do
depend on the values of the order parameters 
$\{m^0,\bar m^0,\Deltad^0,\Deltac^0\}$ at the input layer.

Depending on the structure parameter $\rho$, there are intervals
of the load $\alpha$ where the numbers $\mathcal N_\mathrm{s}$ and
$\mathcal N_\mathrm{u}$ of stable and unstable fixed points
range from 0 to 4.
In particular, for $\rho>\rho_\times$,
\begin{equation*}
\begin{split}
&\left(\mathcal N_\mathrm{s},\mathcal N_\mathrm{u}\right)
= (3,2), (4,3), (2,1), (1,0)\\
&\mathrm{for}\: \alpha\in 
(0,\alpha_\mathrm{p}), (\alpha_\mathrm{p},\alpha_\mathrm{r}),
(\alpha_\mathrm{r},\alpha_\mathrm{c}), (\alpha_\mathrm{c},\infty);
\end{split}
\end{equation*}
for $\rho_\mathrm{c}<\rho<\rho_\times$,
\begin{equation*}
\begin{split}
&\left(\mathcal N_\mathrm{s},\mathcal N_\mathrm{u}\right)
= (3,2), (4,3), (3,2), (1,0)\\
&\mathrm{for}\: \alpha\in 
(0,\alpha_\mathrm{p}), (\alpha_\mathrm{p},\alpha_\mathrm{c}),
(\alpha_\mathrm{c},\alpha_\mathrm{r}), (\alpha_\mathrm{r},\infty);
\end{split}
\end{equation*}
and for $\rho<\rho_\mathrm{c}$,
\begin{equation*}
\begin{split}
&\left(\mathcal N_\mathrm{s},\mathcal N_\mathrm{u}\right)
= (3,2), (3,4), (3,2), (1,0)\\
&\mathrm{for}\: \alpha\in 
(0,\alpha_\mathrm{p}), (\alpha_\mathrm{p},\alpha_\mathrm{c}),
(\alpha_\mathrm{c},\alpha_\mathrm{r}), (\alpha_\mathrm{r},\infty).
\end{split}
\end{equation*}
One always has that $\left|\mathcal N_\mathrm{s} - \mathcal N_\mathrm{u}\right| = 1$.

Which one of the stable fixed points the dynamics converges to
is determined by the initial condition.
The set of initial conditions that converge to a given fixed point
is that fixed point's attraction basin.
The initial conditions, at fixed $\alpha$ and $\rho$, are parameterized by
the two order parameters $m^0$, $\bar m^0$,
because $\Deltad^0$ and $\Deltac^0$ are fixed by Eq.~(\ref{eq:delta_initial}).
Therefore, attraction basins in our model are generically 2-dimensional
at fixed $\rho,\alpha$.
The full phase diagram would be a 4-dimensional plot of the fixed point
as a function of $\{\rho,\alpha,m^0,\bar m^0\}$.
Figure \ref{fig:basins} shows a choice of 2-dimensional sections of the full
phase diagram, where $\rho$ and $\bar m^0$ are fixed,
and $\alpha$ and $m^0$ are varied.
The critical loads $\alpha_\mathrm{c},\alpha_\mathrm{r},\alpha_\mathrm{p}$
emerge clearly in these plots, coherently with their characterizations
in Eq.~(\ref{eq:def_alphas}).

Note that, given a value $\rho$ of the structure parameter,
$m^0$ and $\bar m^0$ must satisfy geometric constraints.
For instance, if $m^0=1$ then the input $\xi$ coincides with $\xi^0_1$,
which means that its overlap $\bar m^0$ with $\bar\xi^0_1$ must be equal to $\rho$.
Let us consider the Hamming distance $d_\mathrm{H}$ between the three strings 
$S^0,\xi^0_1,\bar\xi^0_1\in\{-1,1\}^N$.
The overlap between two strings with Hamming distance $d_\mathrm{H}$
is $1-2d_\mathrm{H}/N$.
It is then easy to see that
\begin{equation}\label{eq:constraints}
\begin{split}
&\rho-1\leq m^0-\bar m^0\leq 1-\rho\\
&m^0+\bar m^0\leq 1+\rho\\
&m^0+\bar m^0\geq -1-\rho\\
\end{split}
\end{equation}
The first 3 inequalities are consequences of the triangle inequality,
and the last one is due to the fact that
$d_\mathrm{H}(S^0,\xi^0_1)+d_\mathrm{H}(\xi^0_1,\bar\xi^0_1)+d_\mathrm{H}(\bar\xi^0_1,S^0) \leq 2N$.
The intervals denoted ``realizable'' in Fig.~\ref{fig:basins}
are those for which all constraints in (\ref{eq:constraints}) are satisfied.
We checked that all fixed points in the bifurcation diagrams, for all values of $\rho$,
satisfy the constraints: there are no
``non-physical'' solutions to the fixed point equations.


\section{Discussion}

The point $\rho_\times$ has an interesting interpretation in terms of the network's
capacity to identify a given pair of patterns.
Let us define the identification capacity $\alpha_\mathrm{i}(\rho)$
as the largest load such that $S^L$, the representation at layer $L$ of the input $S^0$,
has a non-zero overlap with at least one of the patterns in the pair $\{\xi^L_1,\bar\xi^L_1\}$,
in the infinitely deep limit $L\to\infty$:
\begin{equation}
\alpha_\mathrm{i}(\rho):=\sup \left\{ \alpha \,|\, \mathcal S(\alpha,\rho)
\neq \mathcal S_0(\alpha, \rho) \right\}.
\end{equation}
By using the fact that 
$S(\alpha,\rho) = S_\mathrm{r}(\alpha,\rho) \cup S_\mathrm{\bar r}(\alpha,\rho)$,
one has
\begin{equation*}
\begin{split}
\alpha_\mathrm{i}(\rho)&=\inf \left\{ \alpha \,|\, \mathcal S(\alpha,\rho)
= \mathcal S_0(\alpha, \rho) \right\} \\
&=\inf \left\{ \alpha \,|\, S_\mathrm{r}(\alpha,\rho) = \emptyset
\land \mathcal S_\mathrm{\bar r}(\alpha,\rho) = \mathcal S_0(\alpha, \rho) 
\right\}\\
&=\max \left\{ \alpha_\mathrm{c}(\alpha,\rho), \alpha_\mathrm{r}(\alpha,\rho) \right\}.
\end{split}
\end{equation*}
Therefore, the identification capacity is a non-monotonic function of $\rho$,
with a minimum at $\rho_\times$.
This point identifies, within the parameterized model of data structure,
the worst performance of the network in terms of identification capacity.

The non-monotonicity of $\alpha_\mathrm{i}(\rho)$ 
as a function of $\rho$ has another remarkable consequence.
Imagine starting at the unstructured point $\rho=0$. 
Increasing $\rho$ brings patterns in the same pair closer to each other.
Initially, this has a detrimental effect on the identification capacity, due to the correlations introduced.
However, when the structure parameter exceeds a threshold
$\rho_*=0.36$, then one has $\alpha_\mathrm{i}(\rho)>\alpha_\mathrm{i}(0)$,
and $S^L$, at large $L$, maintains a finite overlap with at least one of the patterns $\{\xi^L_1,\bar\xi^L_1\}$
for larger loads than in the unstructured $\rho=0$ case.
The value $\rho_*$ then pinpoints the minimum amount of structure
that a data set must have in order for it to be beneficial,
instead of detrimental, to the identification of two-point classes.
This may be understood as the effect of
competition between memorization of correlated patterns within a class
and identification of class-wide features.

Our definition of layer-wise processing in an unsupervised setting
shares similarities with analogous models studied in different contexts.
First, it may be interpreted as a particular ensemble of random neural networks
\cite{LutherSeung:2020,PooleLahiri:2016},
where the disorder is specified indirectly via the patterns $\xi^l_\mu$.
This feature is what allows us to implement meaningful structure in the disorder.
Second, if the layer index $l$ is interpreted as a (physical) discrete time,
then the model is equivalent to classic models of Hebbian sequence memories,
based on Hopfield models with asymmetric interactions
\cite{ChaudhryZavatoneVeth:2023, SompolinskyKanter:1986, GutfreundMezard:1988}.
In this setting, our parameterization of structured disorder corresponds
to storing sequences of patters where different time frames are uncorrelated
but different sequences have correlated frames.

We have fixed a layer-independent structure parameter $\rho$ in our analysis.
An interesting extension of the model,
relevant also in the sequence-learning perspective discussed above,
is obtained by parameterizing
data structure independently at each layer, thus considering
a multi-component structure parameter $\rho^l$.
Such a description would realize a simple and analytically tractable model of
layer-wise geometric information processing by a neural network
\cite{stragglers, AnsuiniLaio:2019}.


\begin{acknowledgments}
We thank Matteo Osella and Francesco Ginelli
for discussions related to this work.
\end{acknowledgments}


\appendix*
\section{Derivation of the recurrence equations}

In this Appendix, it is understood that sums over indices $i,j$ run from $1$ to $N$,
and sums over indices $\mu,\nu$ run from $1$ to $P$;
any additional constraint, such as ``$\mu\neq 1$'', restricts the respective range.

We proceed explicitly with the calculations for \(m'\) (the same applies to \(\bar m'\))
\begin{equation}\label{eqn:m'_1}
 m' = \langle\, M'_1 \,\rangle = \frac{1}{N} \sum_{i} \langle\, S'_i \, \xi'_{i1} \,\rangle = \langle\, S'_i \, \xi'_{i1} \,\rangle
\end{equation}
The right-hand side can be obtained by using the definition (\ref{eqn:law}) with weight matrix given by (\ref{eqn:hebb}).
We can rearrange the terms to recognize \(M_\nu\) and \(\bar M_\nu\) as follows:

\[
\begin{split}
 S'_i \xi'_{i1} &
 = \xi'_{i1} \sgn \left( \sum_{\mu} \xi'_{i\mu} \, M_{\mu} + \sum_{\mu} \bar \xi'_{i\mu} \, \bar M_{\mu} \right).
\end{split}
\]
We can go further by leveraging the fact that
the units can assume only values \( \pm1 \) and the activation function is the sign;
then $\xi'_i \sgn (\cdot) = \sgn (\xi'_i \cdot)$.
Separating the contribution of \(M_1\) and \(\bar M_1\) from the others, we obtain
\begin{equation}\label{eq:Sxi}
\begin{split}
 S_i' \cdot \xi_{i1}'
 = \sgn \biggr[ &M_1 + \xi'_{i1}  \bar \xi'_{i1}  \bar M_1 \\
 &+ \sum_{\mu \neq 1} \Bigl( \xi'_{i1}  \xi'_{i\mu} M_{\mu} + \xi'_{i1}  \bar \xi'_{i\mu}  \bar M_{\mu} \Bigl)  \biggr] 
\end{split}
\end{equation}

The overlap \(M_{\nu}\) (and similarly for \(\bar M_{\nu}\)) is the mean of \(N\) variables \(S_i \, \xi_{i \nu}\)
that are independent and identically distributed, then from the central limit theorem \(M_{\nu} = m_{\nu} + O(1/\sqrt{N})\).
Recall the assumption, discussed in Sec.~\ref{sec:solution}, that the condition
in Eq.~(\ref{eqn:inital_M}) holds at all layers.
In the thermodynamic limit, we can replace \(M_1\) and \(\bar M_1\) with their average values \(m\) and \(\bar m\).
We cannot do the same for $M_\nu$ and $\bar M_\nu$ with $\nu>1$ because they appear within sums
running over $P$ elements, and $P$ is extensive.

To progress, we assume that the term 
\begin{equation}
x := \sum_{\mu \neq 1} \Bigl( \xi'_{i1} \xi'_{i\mu} M_{\mu} + \xi'_{i1} \bar \xi'_{i\mu} \bar M_{\mu} \Bigl)
\end{equation}
is a normal random variable.
By use of 
\(\left< \xi'_{i\nu} \xi'_{i\mu} \right> = \delta_{\nu \mu}\) 
and 
\(\left< \xi'_{i\nu} \bar\xi'_{i\mu} \right> = \rho\,\delta_{\nu \mu}\), 
we can compute the mean 
\(\left< x \right> = 0\)
and variance $\left<x^2\right>=\Delta$, where
\begin{equation}
\Delta
 = \sum_{\mu \neq 1} \left< M_{\mu} M_{\mu} + \bar M_{\mu} \bar M_{\mu} \right> + 2\rho 
 \sum_{\mu \neq 1} \left<M_{\mu}  \bar M_{\mu}\right>.
 \end{equation}

Since the overlap between patterns belonging to the same pair is \(\rho\),
we have that \(\xi'_{i1} \bar \xi'_{i1} = +1 \) with probability \((1+\rho)/2\)
and \(\xi'_{i1} \bar \xi'_{i1} = -1 \) with probability \((1-\rho)/2\).
The right-hand side of Eq.~(\ref{eq:Sxi}) can then be written as an average
over $x$ and $\xi'_{i1}\bar\xi'_{i1}$:
\begin{equation*}
\begin{split}
 m'
 = \int_{-\infty}^{+\infty} dx \, \mathcal{N}_x(0,\Delta) &\left[\frac{1+\rho}{2}\sgn(m+\bar m+x) \right.\\ 
 &+ \left.\frac{1-\rho}{2}\sgn(m-\bar m+x)\right],
\end{split}
\end{equation*}
which can be expressed in terms of the erf function as in
Eq.~(\ref{eqn:recursive_m}).
The recurrence relation for \(\bar m\), Eq.~(\ref{eqn:recursive_barm}), 
is then obtained exchanging terms \(m\) and \(\bar m\) in (\ref{eqn:recursive_m}).

Now, if one tries to compute $\Delta'$, one realizes that the recurrence equations do not
close on the set $\{m,\bar m, \Delta\}$.
Instead, as discussed in the main text, it is natural to introduce two additional order parameters:
\begin{equation}
\begin{split}
\Deltad &= \sum_{\mu\neq 1} \left<M_{\mu}M_{\mu} + \bar M_{\mu}\bar M_{\mu}\right>\\
\Deltac &=  2\rho \sum_{\mu\neq 1} \left<M_{\mu}\bar M_{\mu}\right>.
\end{split}
\end{equation}
In terms of these, the variance of $x$ is $\Delta=\Deltad+\Deltac$.
To compute $\Deltad'$ and $\Deltac'$, one needs to express $\left<M_\mu' M_\mu'\right>$,
$\left<\bar M_\mu' \bar M_\mu'\right>$, and $\left<M_\mu' \bar M_\mu'\right>$
in terms of unprimed variables.
Note that $\mu\neq 1$ hereafter.
We explicitly show the calculation for \(\left< M'_{\mu}\bar M'_{\mu} \right>\); 
the other terms can be found similarly.
Separating the sum between diagonal and off-diagonal contributions,
we get
\begin{equation*}
 \langle\, M'_{\mu} \bar M'_{\mu} \,\rangle
 = \frac{1}{N^2} \left[ \sum_{i} \left< \xi'_{i\mu} \bar \xi'_{i\mu} \right> 
 + \sum_{i \neq j} \left< S'_i \xi'_{i\mu} S'_j \bar \xi'_{j\mu} \right> \right].
\end{equation*}
The diagonal term gives \(N\rho\) because \(\left<\xi_{i\mu}\bar \xi_{i\mu}\right>=\rho\).
Note that, when computing \(\left< M'_{\mu} M'_{\mu} \right>\) or \(\left< \bar M'_{\mu} \bar M'_{\mu} \right>\), 
the diagonal term gives \(N\) instead.

Proceeding with the calculation of the off-diagonal term, we obtain
\begin{equation}
\label{eqn:MM}
 \left< M'_{\mu} \bar M'_{\mu} \right>
 = \frac{\rho}{N} + \frac{1}{N^2} \sum_{i \neq j} \left<\sgn(H'_{i\mu})\,\sgn(K'_{j\mu})\right>,
\end{equation}
where we introduced
\begin{equation}\label{eq:def_H}
\begin{split}
 H'_{i\mu} 
 := \; & \xi'_{i\mu} \xi'_{i1} m
 + \xi'_{i\mu} \bar \xi'_{i1} \bar m
 + M_{\mu} 
 + \xi'_{i\mu} \bar \xi'_{i\mu} \bar M_{\mu}\\
 &+ \sum_{\nu \neq 1,\mu} \Bigl( \xi'_{i\mu} \xi'_{i\nu} M_{\nu} + \xi'_{i\mu} \bar \xi'_{i\nu} \bar M_{\nu} \Bigl)
\end{split}
\end{equation}
\begin{equation}\label{eq:def_K}
\begin{split}
 K'_{j\mu}
 := \; & \bar \xi'_{j\mu} \xi'_{j1} m
 + \bar \xi'_{j\mu} \bar \xi'_{j1} \bar m
 + \bar M_{\mu}
 + \xi'_{j\mu} \bar \xi'_{j\mu} M_{\mu}\\
&+ \sum_{\nu \neq 1,\mu} \Bigl( \bar \xi'_{j\mu} \xi'_{j\nu} \, M_{\nu} + \bar \xi'_{j\mu} \, \bar \xi'_{j\nu} \, \bar M_{\nu} \Bigl)
\end{split}
\end{equation}
where, as before, we replaced \(M_1\) and \(\bar M_1\) with \(m\) and \(\bar m\),
but kept the random variables in the extensive sums.

The averages in Eq.~(\ref{eqn:MM}) can be decomposed into two averages
over independent ensembles: one on the primed variables $\xi'_{**}, \bar\xi'_{**}$,
and the other on the unprimed variables $M_*, \bar M_*$.
Let us denote the average on the first ensemble as $\left<\cdot\right>_{\xi'}$
and that on the second as $\left<\cdot\right>_M$.
Then, considering that variables with different indices $i,j$
are independent in Eqs.~(\ref{eq:def_H}) and (\ref{eq:def_K}), one has
\begin{equation*}
\begin{split}
\left< \sgn(H'_{i\mu})\sgn(K'_{j\mu})\right> &=
\left<\left<\sgn(H'_{i\mu})\sgn(K'_{j\mu})\right>_{\xi'}\right>_M\\
& = \left<\left<\sgn(H'_{i\mu})\right>_{\xi'} \left<\sgn(K'_{j\mu})\right>_{\xi'}\right>_M
\end{split}
\end{equation*}

We make the assumption that the random variable
\begin{equation}
z := \sum_{\nu \neq 1,\mu} \xi'_{i\mu} \xi'_{i\nu} \, M_{\nu} + \xi'_{i\mu} \bar \xi'_{i\nu} \bar M_{\nu} 
\end{equation}
is normal. Its mean is $\left<z\right>=0$ and its variance is, in the thermodynamic limit,
$\left< z^2\right>=\left<x^2\right>=\Delta$.
Let us define 
\begin{equation}
\eta:=\xi'_{i\mu}\xi'_{i 1}.
\end{equation}
Since the overlap between patterns in the same pair is \(\rho\),
one has that
\begin{equation}\label{eq:prob_xixi}
\begin{split}
&\mathrm{Prob}\left[\eta = \pm1\right]=\frac{1}{2}\\
&\mathrm{Prob}\left[\xi'_{i\mu} \bar \xi'_{i\mu} = \pm1\right]=\frac{1\pm\rho}{2}\\
&\mathrm{Prob}\left[\eta = \pm \xi'_{i\mu} \bar \xi'_{i 1} \right]=\frac{1\pm\rho}{2}.
\end{split}
\end{equation}
Now, the variance of $z$, in the ensemble where 
$M_*$ and $\bar M_*$ are kept fixed, is self averaging,
therefore we substitute the average variance $\Delta$ in the thermodynamic limit.
Then Eqs.~(\ref{eq:prob_xixi}) allow us to express, as an average over
the random variables $z$, $\eta$, $\xi'_{i\mu}\bar\xi'_{i 1}$, and $\xi'_{i\mu}\bar\xi'_{i\mu}$,
\begin{equation}
\left< \sgn(H'_{i\mu}) \right>_{\xi'} = \int_{-\infty}^{+\infty} dz\, \mathcal{N}_z(0,\Delta) 
\frac{1}{2} \sum_{\eta = \pm 1} \Phi(\eta,z)
\end{equation}
where
\[
\begin{split}
 \Phi(\eta,z)
 = &\frac{(1+\rho)^2}{4}\sgn( \eta m + \eta \bar m + M_{\mu} + \bar M_{\mu} + z ) \\ &
+  \frac{(1-\rho)^2}{4}\sgn( \eta m - \eta \bar m + M_{\mu} - \bar M_{\mu} + z ) \\ &
+ \frac{1-\rho^2}{4}\,\sgn( \eta m - \eta \bar m + M_{\mu} + \bar M_{\mu} + z ) \\ &
+ \frac{1-\rho^2}{4}\,\sgn( \eta m + \eta \bar m + M_{\mu} - \bar M_{\mu} + z ).
\end{split}
\]
Using the identity 
\begin{equation*}
\begin{split}
\int_{-\infty}^{+\infty} dz\, \mathcal{N}_z(0,\Delta) \left[ \sgn(a+b+z) + \sgn(-a+b+z) \right] \\
= 2 \int_{a-b}^{a+b} dz \, \mathcal{N}_z(0,\Delta)
\end{split}
\end{equation*}
and considering the thermodynamic limit, we obtain
\begin{equation*}
\begin{split}
\left< \sgn(H'_{i\mu}) \right>_{\xi'}
 = &\frac{2}{\sqrt{2\pi\Delta}} \left[\frac{1+\rho}{2}\,\operatorname{exp}(-m_+^2/2\Delta) \right.\\
 &\left.+\frac{1-\rho}{2}\,\operatorname{exp}(-m_-^2/2\Delta)\right] (M_{\mu}+\rho \bar M_{\mu}),
\end{split}
\end{equation*}
where \(m_\pm:=m_1 \pm \bar m_1\).
The result for \(\left<\sgn(K'_{j\mu}) \right>_{\xi'} \) can be obtained from this one
simply by swapping \(M_{\mu}\) and \(\bar M_{\mu}\).

By evaluating the full average in Eq.~(\ref{eqn:MM}), and the sum over $i\neq j$ in the thermodynamic limit,
we obtain
\begin{equation*}
\begin{split}
&\langle\, M'_{\mu} \, \bar M'_{\mu} \,\rangle = \\
&\frac{\rho}{N} + \frac{2}{\pi} \frac{\Omega_\rho(m,\bar m,\Delta)}{\Deltad+\Deltac}
 \left<(M_{\mu}+\rho \bar M_{\mu})(\bar M_{\mu}+\rho M_{\mu})\right>
\end{split}
\end{equation*}
with $\Omega_\rho(m,\bar m,\Delta)$ defined as in (\ref{eq:omega}).
A very similar computation gives
 \begin{equation*}
 \begin{split}
  \langle\, M'_{\mu} \, M'_{\mu} \,\rangle
 & = \frac{1}{N} + \frac{2}{\pi} \frac{\Omega_\rho(m,\bar m,\Delta)}{\Deltad+\Deltac}
  \left<(M_{\mu}+\rho \bar M_{\mu})^2\right>  \\
  \langle\, \bar M'_{\mu} \, \bar M'_{\mu} \,\rangle
  & = \frac{1}{N} + \frac{2}{\pi} \frac{\Omega_\rho(m,\bar m,\Delta)}{\Deltad+\Deltac}
  \left<(\bar M_{\mu}+\rho M_{\mu})^2\right>
\end{split}
\end{equation*}
With these expressions, one readily obtains Eqs.~(\ref{eqn:recursive_deltad}) and (\ref{eqn:recursive_deltac}).

\bibliographystyle{unsrt}
\bibliography{biblio}

\end{document}